\documentclass[aps,preprint,
]{revtex4}
\usepackage{pictex,graphicx,color,rotating}     

\long\def\beginnote#1\endnote{\begingroup\ttfamily\small#1\endgroup}
\long\def\beginhidden#1\endhidden{}

\long\def\beginnote#1\endnote{\begingroup\ttfamily\small#1\endgroup}

\def\lapproxeq{\lower .7ex\hbox{$\;\stackrel{\textstyle
<}{\sim}\;$}}
\def\gapproxeq{\lower .7ex\hbox{$\;\stackrel{\textstyle
>}{\sim}\;$}}

\begin{document}

\preprint{}
\preprint{BA-01-47}
\vskip 3.0cm

\title{Branes and Inflationary Cosmology}
\author{Bumseok Kyae and Qaisar Shafi}
\affiliation{
Bartol Research Institute, University of Delaware, Newark, DE 19716, USA.  
}
\date{\today}

\begin{abstract}
We discuss a brane based inflationary scenario in which an initially 
non-supersymmetric configuration involving a D4 brane is dynamically 
transformed into a supersymmteric one in a background space-time geometry 
determined by a stack of D6 branes.  
Inflation is realized in the effective four 
dimensional theory and ends after reaching a stable BPS configuration.  
The scalar spectral index turns out to be 0.98.  
Under some simplifying assumption the reheat temperature, Hubble constant and 
the string scale are 
estimated to be of order $10^{8}$ GeV, $10^{12}$ 
GeV and $10^{16}$ GeV, respectively.  
\end{abstract}
\maketitle

\def\p{\partial}
\def\L{\Lambda}

%
The inflationary scenario~\cite{guth} provides an elegant understanding of the 
observed isotropy of the cosmic microwave background radiation as well as 
the observed small deviations from this isotropy.  Implementation of inflation 
in field theory models typically involves a scalar field with a suitably 
chosen potential that yields an appropriate number of e-foldings, the correct 
magnitude of density fluctuations, etc.~\cite{shafi}.    
However, it is not always easy to 
satisfy the various constraints.  For instance, supergravity (SUGRA) 
corrections can often ruin the nice features of an otherwise respectable 
globally supersymmetric model by providing a mass$^2$ term to the inflaton 
field that is comparable to $H^2$, where $H$ denotes the Hubble constant 
during inflation.  

In recent years, new approaches to supersymmetric gauge theories 
have been extensively studied 
using D branes and other extended objects 
in Type II string theory~\cite{hanany}.  
The Hanany-Witten model~\cite{hanany,kuta}, for instance, shows how  
supersymmetric field theory results can be interpreted geometrically 
as special configurations of branes.    
In Fig.~1, 
we present some supersymmetric configurations of branes~\cite{kuta}.  
\begin{figure}[t]
\label{bps}
\begin{center}
\includegraphics[width=100mm]{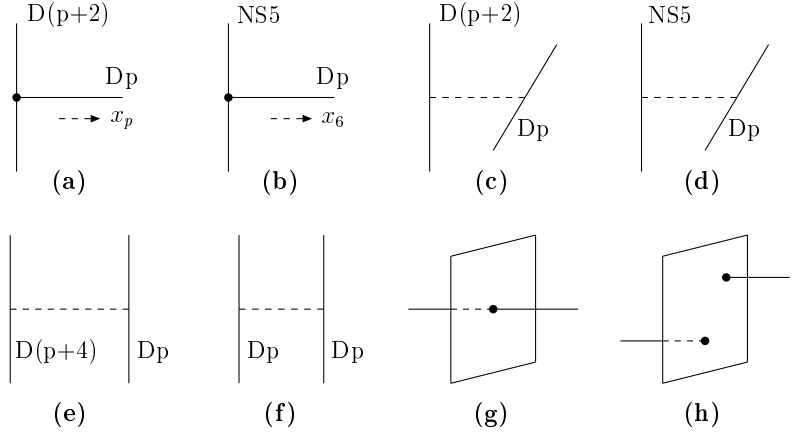}
\end{center}
\caption{
Various supersymmetric configurations of branes.  
${\bf (a)}$ A Dp brane in directions ($x_0, \cdots, x_p$) ends 
in the direction $x_p$ on a D(p+2) brane  
in directions ($x_0, \cdots, x_{p-1},x_{p+1}, x_{p+2}, x_{p+3}$).  
${\bf (b)}$ A Dp brane in directions ($x_0, \cdots, x_{p-1}, x_6$)  
ends in the direction $x_6$ on a NS5 brane in directions ($x_0, \cdots, x_5$).  
${\bf (c)}$ A Dp brane with world volume in directions ($x_0, \cdots, x_p$) and 
a D(p+2) brane 
in directions ($x_0, \cdots, x_{p-1}, x_{p+1}, x_{p+2}, x_{p+3}$).  
${\bf (d)}$ A Dp brane is partially orthogonal to a NS5 brane.  
${\bf (e)}$ A Dp brane in directions ($x_0, \cdots, x_p$) is 
parallel to a D(p+4) brane along ($x_0, \cdots, x_{p+4}$).  
${\bf (f)}$ Two Dp branes are stretched in the same directions.  
In ${\bf (c)}$-- ${\bf (f)}$, dotted lines 
indicate separation between branes.    
While the configurations ${\bf (a)}$--${\bf (g)}$ preserve 8 supercharges, 
${\bf (f)}$ preseves 16 supercharges.  
A Dp brane intersecting a D(p+2) brane as in ${\bf (g)}$ can split into 
two disconnected parts as in ${\bf (h)}$, which separate along the D(p+2) 
brane.  
}
\end{figure}

What then is correspondence in brane picture to inflationary scenario  
in a supersymmetric field theory?  
This is the main focus of our paper.  
To address this, 
it is useful to find a brane setup in which  
an initial non-supersymmetric brane configuration is 
dynamically transformed into 
a supersymmetric one.  
Of course, supersymmetry is broken in nature, although how this breaking 
is realized consistent with an almost vanishing cosmological constant 
remains a fundamental unsolved problem.  
For realizing this, 
D brane--anti-D brane system~\cite{d-inf}, and  
simple Dp--D(p+2) branes and Dp--Dp branes systems~\cite{halyo}  
were previously considered.  
[For earlier pioneering work on brane inflation, see ref.~\cite{dvali}.]    
While a D brane--anti-D brane system is non-supersymmetric 
from the beginning, in \cite{halyo} 
a supersymmetry breaking state was obtained 
through small deviation from a supersymmetric configuration.  
A deviation from configurations in Fig.~1 generically leads 
to supersymmetry breaking  
and a non-trivial potential on a brane, which could be exploited for  
constructing an inflationary model.  
%
%
However, in a simple setup,  
supersymmetry is not restored easily, and inflation does not appear to end.   
An essential reason why this can happen is that 
the induced potential if dominated by the gravitational interaction   
gives rise to an attractive rather than repulsive force, and a mechanism 
must be found to provide a graceful exit from the inflationary phase.  
%
%
The aim of our paper is to construct a consistent model, 
in which an  
initial non-supersymmetric configuration rolls down to a supersymmetric one.  
Our model is inspired by the work on string mediated supersymmetry breaking 
in~\cite{brodie}.  
%
%
\begin{figure}[t]
\label{b1}
\begin{center}
\includegraphics[width=50mm]{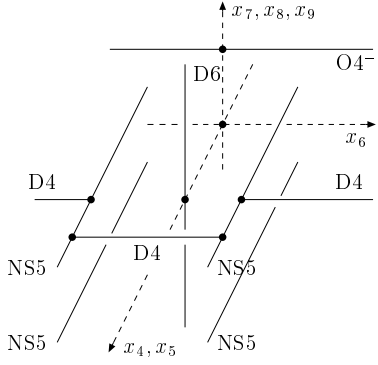}
\end{center}
\caption{Supersymmetric Configuration.
This configuration preserves 8 supercharges.  }
\end{figure}

We consider a configuration 
where a stack of $N$ D6 branes are in 0123789 directiones, with  
four stacks of NS5 branes in 012345 directions as shown 
in Fig.~2 and 
Table 1.    
Let us add D4 branes in 01236 directions as in Fig.~2.  
The configuration preserves eight of thirty two supercharges of 
IIA string theory  
and so is stable.  One can check it usig Fig.~1.  
Additionally, by including an orientifold 4 plane (O$4^-$) 
in the same direction 
as D4 brane, the eight supercharges are still conserved~\cite{kuta}.  
\vskip 0.6cm
\begin{center}
\begin{tabular}{|c||c|c|c|c|c|c|c|c|c|c|} \hline
$x_M$&~~0~~&~~1~~&~~2~~&~~3~~&~~4~~&~~5~~&6(6$'$)&7(7$'$)&~~8~~&~~9~~ \\ 
\hline\hline
D6&N&N&N&N&D&D&D&N&N&N \\ \hline
NS5&$\times$&$\times$&$\times$&$\times$&$\times$
&$\cdot$&$\cdot$&$\cdot$&$\cdot$&$\cdot$ \\ \hline 
D4&N&N&N&N&D&D&N&D&D&D \\ \hline 
O$4^-$&$\times$&$\times$&$\times$&$\times$&
$\cdot$&$\cdot$&$\times$&$\cdot$&$\cdot$&$\cdot$ \\ \hline
%
%
%
\end{tabular}
\vskip 0.2cm
\end{center}
{\bf Table 1.~} Branes and O$4^-$ plane configuration in our model.  
N and D donote Neumann and Dirichlet boundary conditions,  
and `$\times$'s indicate 
the directions in which the corresponding objects stretch.    
\vskip 0.6cm

To drive inflation, we assume that a D4 brane is rotated around 
$x_4$ and/or $x_5$ axes by $\theta$ 
on the $x_6-x_7$ plane 
as shown in Fig.~3.   
The configuration is no longer stable and  
a non-flat potential between D6 and D4 brane is generated.     
Actually, rotation of the D4 brane on the $x_6-x_7$ plane corresponds 
to the generation of Fayet-Iliopoulos term in 
the field theory on the brane~\cite{brodie}.  
Note that the D4 brane--NS5 branes configuration is still BPS.  
%
\begin{figure}[t]
\label{b2}
\begin{center}
\includegraphics[width=100mm]{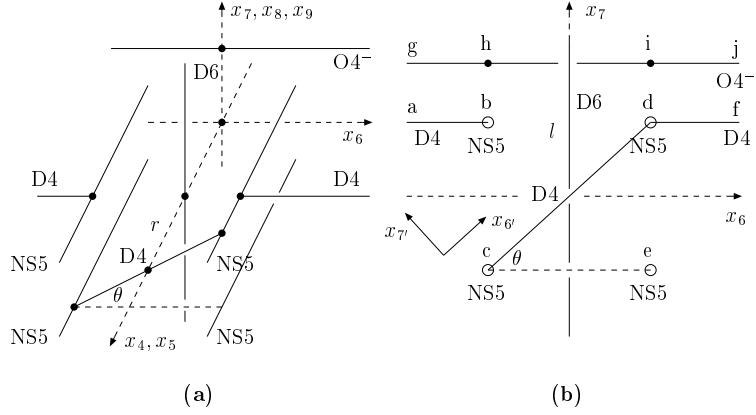}
\end{center}
\caption{Initial Configuration.  
${\bf (a)}$ Non-zero $\theta$ breaks supersymmetry.  
$r$ is the distance between D6 and D4 branes.
${\bf (b)}$ shows the same configuration as ${\bf (a)}$. 
$x_{6'}$ and $x_{7'}$ are the coordinates $x_6$ and $x_7$ rotated   
by $\theta$ around $x_4$ and/or $x_5$.  
An O$4^-$ plane is located on $x_7=l$.    
In toroidal compactification, a--f and 
g--j are identified.   }   
%
\end{figure}

Since the Dp$'$ brane is a source of (p$'$+1) form R-R field  
in Type II string theory, solutions of Dp$'$ branes 
coupled to R-R field can be obtained in the supergravity regime, 
where Type II superstring theory provides us 
with the low energy effective action,  
\begin{eqnarray}\label{sugra}
S_0=\frac{M_s^8}{(2\pi)^7}\int d^{10}x\sqrt{-G}\bigg[
e^{-2\phi}\bigg(R+4(\nabla\phi)^2-\frac{1}{2\cdot 3!}H^2\bigg)
-\frac{2}{(8-p')!}F_{p'+2}^2+\cdots\bigg] ~.   
\end{eqnarray}
Here $M_s \equiv 1 /\sqrt{\alpha'}$, 
$H=dB$ is the 3 form NS-NS field strength, and 
$F_{p'+2}$ is the ($\rm p'+2$) form R-R field strength.  
An extremal solution describing $N$ coincident ${\rm Dp'}$ 
(${\rm p'<7}$) branes 
in string frame~\cite{horowitz} is 
\begin{eqnarray}
ds^2&=&f^{-1/2}(-dt^2+dx_1^2+\cdots+dx_{p'}^2)
+f^{1/2}(dx_{p'+1}^2+\cdots+dx_9^2) ~, \label{graviton}
\\
e^{\phi}&=&g_sf^{(3-p')/4} ~, \label{dilaton}\\
A_{0\dots p'}&=&-\frac{1}{2}(f^{-1}-1) ~, \label{rr}
\end{eqnarray}
where the contribution of $H$ to the solution is ignored.  
$f$ is a harmonic function in the space transverse to the p$'$-brane, 
\begin{eqnarray}\label{warp}
f(r)=1+\frac{d_{p'}g_sN}{(M_s r)^{7-p'}} ~, 
\end{eqnarray}
with 
\begin{eqnarray}
d_{p'}=2^{5-p'}\pi^{(5-p')/2}\Gamma\bigg(\frac{7-p'}{2}\bigg) ~.  
\end{eqnarray}
The solution preserves 16 supercharges.  
%

In our setup, p$'=6$ with N D6 branes in directions 0123789.  
Additionally, there exist NS5 branes in directions 012345, which 
results in the breaking of half the supersymmetries~\cite{kuta}. 
Since they are in BPS state, the setup is stable.  
NS5 brane is the source of the antisymmetric 2 form field $B_{MN}$ 
in the NS-NS sector.  
%
%
A solution of $N_5$ coincident NS5 branes in directions 012345 
in 10 dimensional space-time is given by~\cite{harvey}  
\begin{eqnarray}
ds^2&=&(-dt^2+dx_1^2+\cdots +dx_5^2)+\frac{e^{\phi}}{g_s}
(dx_6^2+\cdots +dx_9^2) ~, \label{ngraviton}\\
e^{\phi}&=&g_s\bigg(1+\sum_j^{N_5}
\frac{1}{M_s^2 (\tilde{r}-\tilde{r}_j)^2}\bigg) ~, \label{ndilaton}\\
H_{IJK}&=&-\epsilon_{IJKM}\partial^M\phi~, \label{ns2}
\end{eqnarray}
where $I, J, K, M$ label the space 
$x_6, \cdots, x_9$ transverse to NS5 branes, and   
$\tilde{r}_j$ denote the positions of NS5 branes.  
This solution preserve 16 supercharges.  
Note that D6 and NS5 branes intersect in 4 dimensional space-time, and  
the resultant effective 4 dimensional space-time is flat.     

For $N$, $N_5>>1$ in Eq.~(\ref{warp}), the above solutions 
(\ref{graviton})--(\ref{rr}) (or (\ref{ngraviton})--(\ref{ns2}))  
could be expected to dominate the background. 
Let us now introduce some `probe' D4 branes    
near a stack of $N$ D6 branes 
in a configuration involving $\theta$ as shown in Fig.~3.  
%
%
With supersymmetry broken, the effect appears in the brane 
action as a non-trivial potential for a `scalar' field.    
Since the D4 branes are perpendicular to NS5 branes,  
the supersymmetric relation with NS5 brane is maintained, and
in the first approximation no potential is generated from this sector.  
We regard (SUSY breaking) backreaction effect on the background
by a `probe' brane as being small.
Thus, we will only consider the potential triggered by D6 branes
which are not in supersymmetric relation with the D4 brane.
%

The low energy dynamics of the world volume of 
Dp brane is generally 
described by the Dirac-Born-Infeld (DBI) action,   
\begin{eqnarray} \label{dbi}
S_{DBI}=-T_p\int d^{p+1}\sigma ~e^{-\phi}
\sqrt{-{\rm det}(g_{ab}+B_{ab}+2\pi\alpha'F_{ab})} ~,  
\end{eqnarray}
where 
\begin{eqnarray}
g_{ab}(\sigma)&=&
\frac{\partial X^M}{\partial\sigma^a}\frac{\partial X^N}{\partial\sigma^b}
~G_{MN}(X(\sigma)) ~,  \label{indmetric}\\
B_{ab}(\sigma)&=&
\frac{\partial X^M}{\partial\sigma^a}\frac{\partial X^N}{\partial\sigma^b}
~B_{MN}(X(\sigma)) ~,  \label{indanti}
\end{eqnarray}
are the induced metric and antisymmetric tensor on the brane\footnote{
The coupling of D brane to background R-R fields is given by the Wess-Zumino 
term, $S_{WZ}=T_p\int A\wedge exp(B+2\pi \alpha'F)|_{\rm p+1~ form}$, 
where $A$'s are R-R fields. 
In our paper, however, we neglect it because it is not relevant to our 
backgroud solution.  } .  
$F_{ab}$ is the field strength of gauge field living on Dp brane.  
[$B_{ab}$ and $F_{ab}$ will not contribute to the leading term  
in the Dp brane potential.]   
The DBI action together with its fermionic part
composes the super-Yang-Mill action.
Hence, it corresponds to leading approximation in supergravity theory.  
The p-brane tension or R-R charge, $T_p$ is determined by perturbative 
string calculation~\cite{pol},  
\begin{eqnarray} \label{rrcharge}
T_p=\frac{M_s^{p+1}}{(2\pi)^pg_s}~.  
\end{eqnarray} 

When a D4 brane is placed near a stack of D6 branes as in Fig.~3,  
its action is derived  
using Eq.~(\ref{graviton}), (\ref{dilaton}) 
and (\ref{indmetric}),   
\begin{eqnarray} \label{D4action}
S_{D4}&\approx&-\frac{T_4}{g_s}\int d^4x\int dx_{6'}\bigg[
({\rm cos}^2\theta+f^{-1}{\rm sin}^2\theta)^{1/2}
+\frac{1}{2}\sum_{i}\partial_a X^i\partial^a X^i
\bigg]\nonumber \\
&\approx&-\frac{T_4}{g_s}\int d^4x\int dx_{6'}\bigg[
1-\frac{g_sN}{4}\frac{{\rm sin}^2\theta}{M_s ~r}
+\frac{1}{2}\sum_{i}\partial_a X^i\partial^a X^i
\bigg]~, 
\end{eqnarray}
where $a=0, 1, 2, 3, 6'$, $i=4, 5, 7', 8, 9$, $r^2= x_4^2+x_5^2$.  
$x_{6'}$ and $x_{7'}$ are the coordinates $x_6-x_7$  
rotated around $x_4$ and/or $x_5$ axes,  
\begin{eqnarray}
x_{6'}&=&x_{6}~{\rm cos}\theta+x_{7}~{\rm sin}\theta \\
x_{7'}&=&-x_{6}~{\rm sin}\theta+x_{7}~{\rm cos}\theta~.      
\end{eqnarray}
$X^i$ denote the transverse fluctuation of D4 brane, and 
$\langle X^i\rangle$ parametrize its location.  
Without loss of generality, we can set $r\equiv \langle X^4\rangle +X^4$.  
When we derive the DBI action for D4 brane in Eq.~(\ref{D4action}), 
we assume that 
\begin{eqnarray} \label{const1}
f_{p'=6}(r)-1=\frac{g_s N}{2M_s r}<1~,   
\end{eqnarray}
in the regime where supergravity
approximation is valid.  
Note that in the following we have assumed that the distance $r$ 
exceeds the size of the D4 brane along the $6'$ direction, so that the D4 brane
`c--d' is essentially `point-like' in the three transverse direction.

A non-zero angle $\theta$ generates a non-trivial attractive 
potential,  
\begin{eqnarray}\label{5dpot}
V_{5d}=\frac{T_4}{g_s}\bigg(1-\frac{g_sN}{4}
\frac{{\rm sin}^2\theta}{M_s ~r}\bigg) 
=\frac{M_s^5}{(2\pi)^4g_s^2}\bigg(1-\frac{g_sN}{4}
\frac{{\rm sin}^2\theta}{M_s ~r}\bigg) ~.   
\end{eqnarray}
We note here that the potential shows $-1/r$ behavior because of the 
gravitational interaction in 3 dimensional space transverse to D6 branes.   
We can check  
that the $-1/r$ dependence of the potential in Eq.~(\ref{5dpot})
is generalized to $-1/r^{(7-p')}$ under a Dp$'$ background.
However, D4 branes' potential contributions in a--b and d--f regions 
in Fig.~3--(b) have only  
the first term in Eq.~(\ref{5dpot}) because $\theta =0$.   

From Eq.~(\ref{5dpot}), the motivation why we consider a 
D6-D4 system comes out nicely.  
A non-zero $\theta$ induces an attractive force.  
This behavior must be maintained even as $r\rightarrow 0$ 
where the supergravity approximation certainly breaks down.  
The presence of a minimum at $r\neq 0$ with $\theta \neq 0$ 
would imply 
the existence of stable non-BPS state, 
which makes no sense.    
Fig.~1, 
presents some examples of BPS stable 
supersymmetric configurations~\cite{kuta}.  
As the Dp, Dp$'$ branes collide,   
we require that  
the system becomes supersymmetric 
in order to make inflation end successfully.  
Thus, the case ${\rm p'=p+2}$ looks promising, 
but only if they can be orthogonal when they meet. 
Actually, this turns out to be true, as we will see.  

The rotation of Dp branes with p$<4$ in extra dimensional space is not well
defined unless we give up the isotropy of our three space, 1,2,3.
Besides, in Eq.~(\ref{warp}), ${\rm p'}$ should be smaller than 7.  
Therefore, the case that meets these requirements corresponds to 
${\rm p'=6}$ and ${\rm p=4}$.   

In Eq.~(\ref{5dpot}), 
the first term is the tension or R-R charge of D4 brane,   
which contributes to the cosmological 
constant of the effective 4 dimensional theory. 
An economical way to compensate the flux from the  
positive R-R charge of D4 brane in compact space is  
to introduce an orientifold 4 (O$4^-$) plane 
carrying negative R-R charges as in Fig.~3.  
Since it is impossible to separate two NS5 branes along the orientifold
when the charge of the orientifold is negative~\cite{kuta},  
we set a O$4^-$ plane apart from NS5 and D6 brane.    

An O$4^-$ plane in directions 01236, does not spoil  
any remaining supersymmetry when $\theta=0$.  
Recall that 
the O$4^-$ plane is a generalization of $Z_2$ orbifold fixed plane to 
non-oriented string theories and identifies  
$x_k$ and $-x_k$ for $k=4, 5, 8, 9$, and $x_7-l$ and $-x_7+l$.     
In our case, since we assume toroidal compactification for
the extra dimesions $x_4, \cdots, x_9$,  
there exist $2^5$ O$4^-$ planes.  
Thus, there exist $2^5-1$ more `image' setups as shown in Fig.~3  
as well as the original setup.  
Since the R-R charge of Op$^-$ plane is 
$Q_{Op^-}=-2^{p-4}Q_{Dp}$~\cite{kuta},  
the number of R-R charges of D4 branes 
and O$4^-$ planes are the same in 10 dimensional space-time.      
However,  
in the effective 4 dimensional theory obtained 
after integrating out extra dimensions such as $x_6$ or $x_{6'}$, 
the cosmological constant 
does not vanish, 
\begin{eqnarray} \label{effrr}
T_4\int_c^d dx_{6'}-T_4\int_h^i dx_{6}
=T_4~R_{6'}(1-{\rm cos}\theta)~>0~, 
\end{eqnarray}
where $R_{6'}$ denotes the length of D4 brane `c--d', 
\begin{eqnarray}
R_{6'}\equiv \int_c^ddx_{6'}~.
\end{eqnarray} 

The full effective 4 dimensional potential is given by 
\begin{eqnarray} \label{4dpot}
V_{4d}&=&2{\rm sin}^2\frac{\theta}{2}\frac{M_s^5R_{6'}}{(2\pi)^4g_s^2}
\left[1-\frac{N{\rm cos}^2\frac{\theta}{2}}{2(2\pi)^2}\cdot 
\frac{M_s^{3/2}R_6^{1/2}}{\phi}\right]\nonumber \\
&\equiv&M^4\bigg(1-\frac{m}{\phi}\bigg)~,  
\end{eqnarray}
using Eq.~(\ref{rrcharge}) and (\ref{effrr}).  
$\phi$ is the canoically normalized scalar field proportional to 
$r$ $(=\langle X^4\rangle +X^4)$,    
\begin{eqnarray} \label{phi}
\phi=\frac{M_s^{5/2}R_{6'}^{1/2}}{(2\pi)^2g_s}\times r~.   
\end{eqnarray}
Thus, the potential vanishes if the system is in the BPS state,  
namely, $\theta =0$.  
The first term in the above potential corresponds to the Fayet-Iliopoulos 
D-term coefficient $g_{SYM}^2\xi^2$ 
in the 4 dimensional field theory Lagrangian on the brane~\cite{brodie}.  

The above potential is positive definite 
only if Eq.~(\ref{const1}) is satisfied,  
in the regime where the supergravity   
approximation is valid.    
We intend to implement inflation with the above potential.  
Indeed, our inflationary scheme basically corresponds to 
a D-term inflationary model in field theory.  
To end inflation,  
the non-supersymmetric configuration should roll into a BPS configuration,  
which would remove the non-trivial $\phi$ dependence of the potential, and 
the length of D4 brane `c--d' in Fig.~3 should become the same as `h--i', 
which would cancel the first term in Eq.~(\ref{4dpot}).      

Let us now explain   
the roles played by NS5 branes.  
They help to make the length of the `tilted' D4 brane finite, 
which is a necessary condition in the presence of O$4^-$ plane.  
Unless the length of the `tilted' D4 brane 
with respect to O$4^-$ plane is finite,   
two semi-infinite D4 branes should intersect each other 
due to orbifold symmetry introduced by the O$4^-$ plane, 
which is an undesirable situation.    
Thus, introduction of the NS5 brane `d' in Fig.~3--(b) is necessary.  
Then, to cancel the R-R charge contribution `i--j' of O$4^-$ plane 
in Fig.~3--(b), we should set a D4 brane at `d--f'.     
Since `a' and `f' in Fig.~3--(b) 
are identified in toroidal compactification, 
introduction of the D4 brane `d--f' should be accompanied 
by introduction of NS5 brane `b' 
to prevent the D4 brane from reaching `d'.  
The existence of the NS5 brane `c' makes it possible that the D4 brane `c--f'  
takes a supersymmetric configuration in the final stage.     
The NS5 brane `e' is not indispensable in our model.  
   
The potential has a minimum at the origin, as we argued already.   
In ten dimensional space-time, 
the background D6 branes attract the probe D4 brane.  
However, when $r\approx 1/M_s$, 
supergravity is no longer a good approximation.  
With the tachyon condensation of 4-6 strings, 
the configuration rolls to a supersymmetric Higgs 
branch~\cite{kuta,brodie}.  

As the D6 and D4 branes merge, 
the D4 brane is split, and the end points of the split branes 
are able to move around on the D6 brane\footnote{Under U duality,  
the D6--D4 system is transformed into a system 
in which a fundamental string ends on a D3 brane. 
The string end point is able to move around on the D3 brane.  
See Fig.~1 (g)--(h).  }.   
In fact, ${\rm sin}\theta$ could in principle be a dynamical field 
related to $X^{7'}$,
but in our case it was fixed due to the two NS5 branes.  
After splitting of the D4 brane, ${\rm sin}\theta$ is no longer fixed and
the D4 branes can rotate around the NS5 branes.
When they are linked to 
the D6 brane with $\theta =0$ as shown in Fig.~4,  
which is the minimum length configuration of D4 brane between D6 
and NS5 branes, the potential energy is minimum.  
The loss of D4 brane length compared with Fig.~3 is transferred into
kinetic energy of the D4 branes, namely, bending and oscillation of D4
branes around the configuration in Fig.~4.
As the two split D4 branes ``cool down'',
the bending and oscillation modes would eventually disappear.

The final configuration in Fig.~4 restores $N=2$ supersymmetry~\cite{kuta}, 
which removes the $\phi$ dependence of the potential.   
Additionally, the D4 brane length exactly equals that of the O$4^-$ plane, 
and so the constant term in the potential of the 4 dimensional effective theory 
also vanishes.  
Thus, the universe exits gracefully from the inflationary phase.  
\begin{figure}[t]
\label{b3}
\begin{center}
\includegraphics[width=100mm]{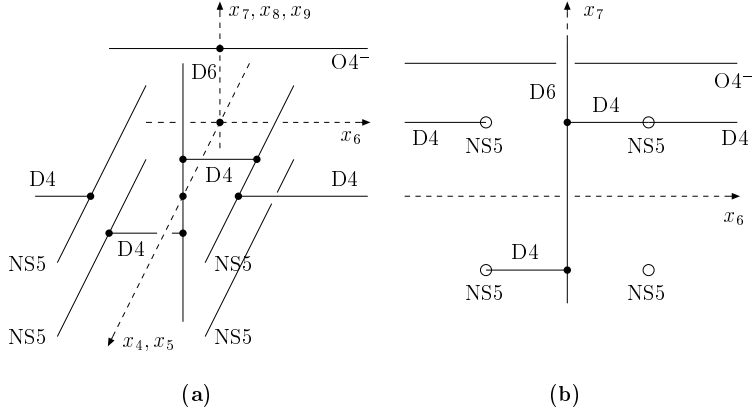}
\end{center}
\caption{Final Configuration.
${\bf (a)}$ D4 brane is split and takes minimum length configuration between 
D6 and NS5 branes.  
At the minimum of the potential, supersymmetry is restored.  
${\bf (b)}$ shows the same configuration as in ${\bf (a)}$.   
}
\end{figure}
%

Let us now discuss some details of inflation 
with the potential (\ref{4dpot}).   
For a desirable inflationary scenario, the flatness conditions on the 
potential should be satisfied, 
\begin{eqnarray}
\epsilon&\equiv& \frac{M_P^2}{2}\left(\frac{V'}{V}\right)^2
\approx \frac{M_P^2}{2}\left(\frac{m}{\phi^2}\right)^2<<1~,  \\
|\eta|&\equiv& |M_P^2\frac{V''}{V}|\approx 2M_P^2\frac{m}{\phi^3}<<1~,  
\end{eqnarray}
where $M_P\approx 2\times 10^{18}$ GeV is the 4 dimensional Planck mass, 
and primes indicate derivatives with respect to $\phi$.   
The critical value of $\phi$ at which inflation ends  
is  
\begin{eqnarray} \label{criticalr}
\phi_c^3\approx 2mM_P^2~.  
\end{eqnarray}
The number of $e$-folds $N_e$ during inflation to solve the smoothness and 
flatness problems is given by 
\begin{eqnarray}
N_e\equiv \frac{1}{M_P^2}\int_{\phi_f}^{\phi_i}d\phi\frac{V}{V'}
\approx \frac{2}{3}\bigg(\frac{\phi_i}{\phi_c}\bigg)^3\approx 60 ~.
\end{eqnarray}
Hence, the initial value of $\phi$ should be $\phi_i\approx 4.5 \phi_c$.  
The scalar spectral index of density fluctuations 
is in excellent agreement with current  
measurements~\cite{spectralindex},  
\begin{eqnarray}
n&\approx&1+2\epsilon -3\eta \nonumber \\
%
&\approx&1-\frac{4}{3N_e}\approx 0.98~.
\end{eqnarray}
From the power spectrum measured by COBE 
at scale $k\approx 7.5H_0$ \cite{cobe}, 
\begin{eqnarray}
\delta_H\equiv\frac{2}{5}{\cal{P_R}}^{\frac{1}{2}}
\equiv\bigg[\frac{1}{75\pi^2M_P^6}\frac{V^3}{V^{'2}}\bigg]^{
\frac{1}{2}}\approx 1.91\times 10^{-5} ~,  
\end{eqnarray}
the length of D4 brane `c--d' in Fig.~3 is constrained to be 
\begin{eqnarray} \label{R6}
R_{6'}\sim 10^{-12}\times \bigg(N{\rm cos}^2\frac{\theta}{2}\bigg)^{-2}
\bigg(\frac{N{\rm cos}^2\frac{\theta}{2}g_s}
{2{\rm sin}\frac{\theta}{2}}\bigg)^3\bigg(\frac{M_P}{M_s}
\bigg)^6\frac{1}{M_P} ~.  
\end{eqnarray}
From Eq.~(\ref{4dpot}) and (\ref{R6}), 
the Hubble constant during inflation is 
\begin{eqnarray} \label{hubble}
H 
~\sim ~10^{-8}\times \bigg(
\frac{N{\rm cos}^2\frac{\theta}{2}g_s}{2{\rm sin}\frac{\theta}{2}}
\bigg)^{1/2}\bigg(\frac{M_P}{M_s}\bigg)^{1/2}M_P ~.
\end{eqnarray}
With Eq.~(\ref{criticalr}), (\ref{phi}), 
(\ref{4dpot}) and (\ref{R6}), the critical value of 
$r$ at which inflation ends is given by 
\begin{eqnarray} \label{rc}
\frac{1}{r_c} \approx 10^{-5}\times\frac{1}{2{\rm sin}\frac{\theta}{2}}~M_P ~.  
\end{eqnarray}
Note that the D4 brane length along the $6'$ direction is less than $r$
during inflation, which implies that 
\begin{eqnarray} \label{pointlike}
\frac{1}{r_c}\lapproxeq \frac{1}{R_{6'}} ~.
\end{eqnarray}

Once the slow roll condition breaks down, the scalar field $\phi$ 
moves rapidly on the Hubble timescale,   
oscillating at the bottom of the potential.  
As mentioned above, this picture would be interpreted as 
the bending and oscillations of D4 brane 
near the final supersymmetric configuration in ten dimensional space-time.  

Such a ``hot'' D4 brane, which could be a hidden brane, 
heats up the bulk 
through gravitational interaction.  
The decay rate of $\phi$ is roughly estimated to be  
\begin{eqnarray}
\Gamma_\phi &\sim& G_N^2\frac{(M^4)^{\frac{5}{4}}}{M_s^6~L_4L_5\cdots L_9} 
\nonumber \\
&\sim&10^{-20}\times \bigg(
\frac{N{\rm cos}^2\frac{\theta}{2} g_s}{2{\rm sin}\frac{\theta}{2}}
\bigg)^{5/4}\bigg(\frac{M_s}{M_P}\bigg)^{3/4}M_P ~,  
\end{eqnarray}
where $G_N\approx M_P^{-2}$ is the 4 dimensional Newtonian constant,    
$L_4, \cdots, L_9$ label the size of the corresponding extra 
dimensions,  
and we use the relation  
\begin{eqnarray} \label{MP}
M_P^2\sim M_s^8\times (L_4L_5\cdots L_9)~.  
\end{eqnarray}
The bending and oscillation of D4 brane are damped out when the Hubble 
time becomes comparable to $\Gamma^{-1}_{\phi}$, and the bulk 
including other branes 
reheat to a temperature,  
\begin{eqnarray}
T_{r}\sim ~0.1\sqrt{\Gamma_\phi M_P} 
~\sim ~10^{-11}\times \bigg(
\frac{N{\rm cos}^2\frac{\theta}{2}g_s}{2{\rm sin}\frac{\theta}{2}}
\bigg)^{5/8} 
\bigg(\frac{M_s}{M_P}\bigg)^{3/8}M_P\equiv 10^{-x}M_P~, 
\end{eqnarray}
where $x$ parametrizes the reheat temperature $T_{r}$.  
The string scale $M_s$ may be parametrized by 
\begin{eqnarray} \label{ms}
\frac{M_s}{M_P}\sim 10^{8(11-x)/3}\times\bigg(
\frac{2{\rm sin}\frac{\theta}{2}}{N{\rm cos}^2\frac{\theta}{2}g_s}\bigg)^{5/3}
~.  
\end{eqnarray}

Let us discuss the phenomenological constraints on parameters 
and predictions in our model.  
We require that $r_c$ in Eq.~(\ref{rc}) 
respects the constraint from Eq.~(\ref{const1}), which gives 
\begin{eqnarray} \label{1}
10^{-5} \times  \bigg( 
\frac{N{\rm cos}^2\frac{\theta}{2}g_s}{2{\rm sin}\frac{\theta}{2}}
\bigg) <\frac{M_s}{M_P} ~.
\end{eqnarray}
The length $R_{6'}$ in Eq.~(\ref{R6}) of D4 brane 
should exceed $1/M_s$ but be smaller than $r_c$, 
\begin{eqnarray} \label{2}
\frac{M_s}{M_P}&\lapproxeq & 10^{-3}\times \bigg(
\frac{N{\rm cos}^2\frac{\theta}{2}g_s}{2{\rm sin}\frac{\theta}{2}}
\bigg)^{5/3}\bigg(N{\rm cos}^2\frac{\theta}{2}g_s\bigg)^{-2/5}~, \\
\frac{M_s}{M_P}&\gapproxeq & \label{4}
10^{-3}\times \frac{1}{(2{\rm sin}\frac{\theta}{2})^{1/6}}
\frac{1}{( N{\rm cos}^2\frac{\theta}{2}g_s)^{1/3}}
\bigg(
\frac{N{\rm cos}^2\frac{\theta}{2}}{2{\rm sin}\frac{\theta}{2}}\bigg)^{1/2}
~.  
\end{eqnarray}	
Also, from Eq.~(\ref{MP}),  
\begin{eqnarray} \label{3}
\frac{M_s}{M_P}<1 ~.  
\end{eqnarray}

With the above constraints, we derive the following results.   
The value of $T_r$  
is around $10^{8}$ GeV.  
Hence, it is possible to avoid the gravitino problem~\cite{kim} in our model.  
%
%
For $N\sim 10$, $g_s\approx 2$, and $\theta\sim 0.1$,  
%
%
the string scale is   
of order $10^{16}$ GeV.  
%
%
%
The size of the D4 brane `c--d' $R_{6'}$ and the Hubble parameter $H$ 
are estimated to be 
%
%
$1/R_{6'}\sim 10^{15}-10^{16}$ GeV and  
$H\sim 10^{12}$ GeV, respectively.     
%
From the estimate for $H$, we conclude that
the tensor contribution to the quadrupole anisotropy can be safely ignored.  
Since $1/r_c\sim 10^{14}$ GeV for $\theta\sim 0.1$, 
we take $1/L_4$, $1/L_5$, $1/L_6\sim 10^{14}$ GeV. 
Then in Eq.~(\ref{MP}), 
$1/L_7$, $1/L_8$, $1/L_9$ are order of $10^{16}$ GeV.     

In conclusion, we have proposed an inflationary model in which inflation is 
associated with the motion of a (hidden) D4 brane  
from a supersymmetry breaking configuration to a supersymmetric one, 
under the influence of a non-trivial background geometry 
 provided by a stack of $N$ D6 branes.    
The supersymmetry breaking configuration 
is parametrized by an angle $\theta$,    
which corresponds to the 
induced Fayet-Iliopoulos D-term coefficient in the effective 4 dimensional 
theory, and results in an attractive potential.  
We found that for consistency the model requires   
NS5 branes and an orientifold 4 plane (O$4^-$)  
in a special configuration.  
The scalar spectral index is 0.98, in excellent agreement 
with the current measurements.  
It remains to be seen how the scenario can be extended to address issues  
such as the nature of dark energy, origin of baryon asymmetry, etc.  
%
%

\acknowledgments
We thank Gia Dvali for very useful discussions and a critical reading of the  
manuscript.  
We also thank Ashoke Sen for a useful discussion.  
The work is partially supported 
by DOE under contract number DE-FG02-91ER40626.  
\vskip 0.3cm 
{\bf Note Added:} After completion of this work we come across the paper by 
C. Herdeiro, S. Hirano, and R. Kallosh (hep-th/0110271), which also uses 
D4, D6 and NS5 branes to realize inflation, but in a different parameter 
regime.    


\end{document}